\documentstyle[multicol,aps,epsf]{revtex}

\title{Algorithm for molecular dynamics simulations of spin liquids}
\author{I. P. Omelyan,$^1$ I. M. Mryglod,$^{1,2}$ and R. Folk$^2$}
\address{$^1$Institute for Condensed Matter Physics,
         1 Svientsitskii Street, UA-79011 Lviv, Ukraine}
\address{$^2$Institute for Theoretical Physics, Linz University,
         A-4040 Linz, Austria}

\date{\today}

\begin{document}
\maketitle

\begin{abstract}

A new symplectic time-reversible algorithm for numerical integration
of the equations of motion in magnetic liquids is proposed. It is
tested and applied to molecular dynamics simulations of a Heisenberg
spin fluid. We show that the algorithm exactly conserves spin lengths
and can be used with much larger time steps than those inherent in
standard predictor-corrector schemes. The results obtained for time
correlation functions demonstrate the evident dynamic interplay
between the liquid and magnetic subsystems.

\vspace{6pt}

\noindent
Pacs numbers: 02.60.Cb; 75.40.Gb; 75.50.Mm; 76.50.+g; 95.75.Pq

\end{abstract}

\vspace{12pt}

\begin{multicols}{2}

Computer experiments remain an important tool for the prediction
and theoretical understanding of various phenomena in magnetic
materials. The methods of Monte-Carlo (MC) and molecular dynamics
(MD) were intensively exploited over the years for the
investigation of phase diagrams, critical phenomena, scaling, and
the dynamic behavior of {\em lattice} systems such as the Ising,
the XY, and the Heisenberg model \cite{EvLan,BunChen,LanKre}.

The necessity to extend these studies to {\em disordered} models
of magnetic liquids was motivated by a great amount of additional
physical properties arising when both spin (orientational) and
liquid (translational) degrees of freedom are taken into account
\cite{Kob,Lom,Tavar,Nijmei,Mryglod,MryglodFolk}. The computer
experiments for such systems have been restricted to MC
simulations \cite{Lom,Nijmei} in which only static quantities
could be calculated. {\em Dynamic} phenomena, in particular, spin
and density relaxations, and the effects connected with the mutual
influence of magnetic and liquid subsystems can be investigated in
MD simulations. Our special interest to this problem was also
stimulated by the results obtained recently for a Heisenberg fluid
within the hydrodynamic theory \cite{MryglodFolk}. One of them was
the prediction that the shape of magnetic dynamic structure factor
can change {\em qualitatively} in comparison with the lattice
model due to the {\em coupling} between the subsystems.

Until now, there have been {\em no} attempts to simulate spin
liquids within the MD approach. This can be explained by the
absence of an efficient MD algorithm for handling the
corresponding equations of motion (EOM). The traditional numerical
methods \cite{Burden} for solving differential equations are
unsuitable because they become highly unstable on time scales used
in MD simulations. As has been well established for pure liquid
systems \cite{Allen,Omelyan}, even standard predictor-corrector
schemes are not efficient because of poor total energy
conservation.

The properties of an acceptable algorithm for long-time observations
over a many-body system should be: stability, accuracy, speed
and ease of implementation. There exists only a small group
of integrators satisfying these criteria. An important one is the
velocity Verlet (VV) algorithm \cite{Swope,Frenkel} which allows
a high accuracy with minimal costs in terms of
time-consuming force evaluations. However, the VV and other similar
schemes \cite{Allen,Tuckerman} were designed to simulate pure liquid
dynamics. In the case of magnetic liquids the situation is more
complicated since the translational positions and momenta are
{\em coupled} with spin orientations in a characteristic way and,
hence, all these dynamical variables must be considered simultaneously.
This requires substantial revision of the liquid dynamic algorithms.

Recently, new algorithms have been devised for spin dynamics
simulations of lattice systems \cite{Krech}. They are based (like
the VV integrator) on the Suzuki-Trotter (ST) decomposition method
and appear to be much more efficient than predictor-corrector
schemes. These algorithms are applicable to spin systems if the
decomposition on two (or several) noninteracting sublattices is
possible. However, they cannot be used for models with arbitrary
spatial spin distributions and, therefore, not for spin liquids.

In this Letter we develop the idea of using ST-like decompositions for
spin liquid dynamics and derive the desired MD algorithm. This allows
quantitative measurement of dynamical structure factors of a Heisenberg
fluid. The main result obtained (reflecting the influence of the liquid
sybsystem on spin dynamics) is the identification of a new propagative
sound-like mode in the spectrum of collective longitudinal spin
excitations.

Consider a classical system composed of $N$ magnetic particles of
mass $m$, described by the Hamiltonian \cite{Nijmei,Mryglod}
\begin{equation}
H = \sum_{i=1}^N \frac{m {{\bf v}_i}^2}{2} + \sum_{i < j}^N
\Big[ \ \Phi(r_{ij}) - J(r_{ij}) \, {\bf s}_i {\mbox{\boldmath
$\cdot$}} {\bf s}_j \Big] \, .
\end{equation}
Here ${\bf r}_i$ and ${\bf v}_i$ are the translational position and
velocity, respectively, of particle $i$ carrying spin ${\bf s}_i$.
The liquid potential is denoted by $\Phi(r_{ij})$, and $J(r_{ij})>0$
is the exchange integral for a pair of spins with
interparticle distance $r_{ij}$. The classical approach treats
${\bf s}_i$ as a three-component continuous vector with a fixed
length for each site $i$. We put for convenience $|{\bf s}_i| =
1$, so that $J$ is measured in energy units.

In order to study the {\em dynamic} properties, the equations of
motion given by ${\rm d} {\mbox{\boldmath $\rho$}}/{\rm d} t =
L{\mbox{\boldmath $\rho$}}(t)$ must be integrated numerically,
where
\begin{eqnarray}
L&=&\sum_{i=1}^N \Big( {\bf v}_i {\mbox{\boldmath $\cdot$}}
\frac{\partial}{\partial {\bf r}_i}+{\bf a}_i {\mbox{\boldmath $\cdot$}}
\frac{\partial}{\partial {\bf v}_i}+[{\mbox{\boldmath $\omega$}}_i
{\mbox{\boldmath $\times$}} {\bf s}_i ] {\mbox{\boldmath $\cdot$}}
\frac{\partial}{\partial {\bf s}_i} \Big) \nonumber \\ &\equiv&
\sum_{i=1}^N \big( L_{{\bf r}_i}+L_{{\bf v}_i}+L_{{\bf s}_i} \big)
\equiv L_{\bf r}+L_{\bf v}+L_{\bf s} \, ,
\end{eqnarray}
is the Liouville operator, i.e., $L {\mbox{\boldmath $\rho$}} =
[{\mbox{\boldmath $\rho$}}, H]$ with $[\, , \,]$ being the Poisson
bracket; ${\mbox{\boldmath $\rho$}} \equiv \{ {\bf r}_i, {\bf
v}_i, {\bf s}_i \}$ denotes the full set of microscopic phase
variables; ${\bf a}_i={\bf f}_i/m$ and ${\mbox{\boldmath
$\omega$}}_i =-{\bf g}_i/{\hbar}$ are the acceleration and local
Larmor frequency, respectively, with ${\bf f}_i=- \sum_{j (j \ne
i)} [{\rm d} \Phi(r_{ij}) /{\rm d} r_{ij}-{\rm d} J(r_{ij})/{\rm
d} r_{ij} \, {\bf s}_i {\mbox{\boldmath $\cdot$}}{\bf s}_j] {\bf
r}_{ij}/r_{ij}$ and ${\bf g}_i = \sum_{j (j \ne i)} J(r_{ij}) {\bf
s}_j$ being the force and internal magnetic field. Note that the
operators $L_{\bf r}$, $L_{\bf v}$ and $L_{\bf s}$ act only on
position, velocity and spin, respectively, and the quantum Poisson
bracket was used  \cite{Mryglod,Krech} to obtain $L_{\bf s}$.

The solutions can be cast in the form ${\mbox{\boldmath $\rho$}}
(t+h)= e^{L h} {\mbox{\boldmath $\rho$}}(t) = e^{(L_{\bf r}+
L_{\bf v}+L_{\bf s})h} {\mbox{\boldmath $\rho$}}(t)$, where $h$ is
the time step. Since the exponential propagator $e^{L h}$ cannot
be evaluated exactly, one introduces some approximations which
take advantage of the smallness of $h$. Assuming for the moment
that spin variables are frozen, i.e. setting $L_{\bf s} \to 0$, we
come to the usual (liquid-like) EOM. They can be solved in a quite
efficient way using the second-order VV integrator
\cite{Swope,Frenkel} which is based on the ST formula ${\rm
e}^{(L_{\bf r}+L_{\bf v}) h}={\rm e}^{L_{\bf v} h/2} {\rm
e}^{L_{\bf r} h} {\rm e}^{L_{\bf v} h/2}+{\cal O}(h^3)$. Taking
into account the fact that this formula is valid for arbitrary two
operators and unfreezing now the spin variables, we obtain
immediately: ${\rm e}^{(L_{\bf r}+L_{\bf v}+L_{\bf s}) h}= {\rm
e}^{L_{\bf v} h/2}{\rm e}^{(L_{\bf r}+L_{\bf s}) h}{\rm e}^{L_{\bf
v} h/2}+{\cal O}(h^3)$, where the sum $L_{\bf r}+L_{\bf s}$ was
treated as one operator. The spin-position subpropagator can
further be decomposed in a similar way, ${\rm e}^{(L_{\bf
r}+L_{\bf s}) h}={\rm e}^{L_{\bf r} h/2} {\rm e}^{L_{\bf s} h}
{\rm e}^{L_{\bf r} h/2}+{\cal O}(h^3)$, resulting in a full
propagation of the form
\begin{equation}
{\mbox{\boldmath $\rho$}}(t+h)=
{\rm e}^{L_{\bf v} \frac{h}{2}}
{\rm e}^{L_{\bf r} \frac{h}{2}}
{\rm e}^{L_{\bf s} h}
{\rm e}^{L_{\bf r} \frac{h}{2}}
{\rm e}^{L_{\bf v} \frac{h}{2}}
{\mbox{\boldmath $\rho$}}(t)+{\cal O}(h^3) \, .
\end{equation}
Note that other decompositions are also possible, but then the
local fields ${\bf g}_i$ and/or forces ${\bf f}_i$ have to be
updated (the most time-consuming operations) more frequently,
which reduces the efficiency of the computations.

The main idea of the decompositions is to obtain subpropagators which
can be evaluated {\em analytically}. It can be shown \cite{Frenkel} that
the position ${\rm e}^{L_{\bf r} \tau}=\prod_i {\rm e}^{L_{{\bf r}_i}
\tau}$ and velocity ${\rm e}^{L_{\bf v} \tau}=\prod_i {\rm e}^{L_{{\bf
v}_i} \tau}$ propagations represent shift operators, namely, ${\rm
e}^{L_{{\bf r}_i} \tau} {{\bf r}_i} = {{\bf r}_i}+{{\bf v}_i} \tau$
and ${\rm e}^{L_{{\bf v}_i} \tau} {{\bf v}_i} = {{\bf v}_i}+{{\bf a}_i}
\tau$. Since the components $L_{\bf r}$ and $L_{\bf v}$ (as well
as $L_{\bf s}$) do not commute, such shifts must be performed in a
rigorous order (as specified by Eq. (3)) and applied to the current
values of ${\bf r}_i$ and ${\bf v}_i$ within the time step.

The spin subdynamics is described in Eq. (3) by the exponential
operator ${\rm e}^{L_{\bf s} h}$. This operator has no simple
explicit form, because the Larmor frequency ${\mbox{\boldmath
$\omega$}}_i$ for each particle depends in general on the
orientations of all other spins of the system. The explicit
solution, nevertheless, may be found as follows. Since all the
partial components $L_{{\bf s}_i}$ do not commute each other, it
is quite natural to find an ST-like decomposition for the whole
set of these operators. This results to the expression
\begin{equation}
{\rm e}^{L_{\bf s} h} = {\rm e}^{L_{{\bf s}_1} \frac{h}{2}} \ldots
{\rm e}^{L_{{\bf s}_{N-1}} \frac{h}{2}} {\rm e}^{L_{{\bf s}_N} h}
{\rm e}^{L_{{\bf s}_{N-1}} \frac{h}{2}} \ldots {\rm e}^{L_{{\bf
s}_1} \frac{h}{2}} \,,
\end{equation}
which constitutes an ST analog for arbitrary number of operators
and is accurate to the same order ${\cal O}(h^3)$ as the terms
already truncated. Again, other ${\cal O}(h^3)$ decompositions may
be introduced. However, only Eq.~(4) will lead to a scheme with
{\em minimal} number of local field recalculations.

The problem is now considerably simplified because, according to
Eq.~(4), each current value of ${\bf s}_i$ is updated spin by spin
at a fixed instantaneous Larmor frequency ${\mbox{\boldmath
$\omega$}}_i$, and this case allows analytical solutions: ${\rm
e}^{L_{{\bf s}_i} \tau} {\bf s}_i(t) = {\bf D}_i(t,\tau) {\bf
s}_i(t)$. Here ${\bf D}_i(t,\tau)= {\bf I}+{\bf W}_i \sin(\omega_i
\tau)+{\bf W}_i^2 (1-\cos(\omega_i \tau))$ denotes an orthonormal
(${\bf D} {\bf D}^+={\bf I}$) matrix of rotation around axis
${\mbox{\boldmath $\omega$}}_i$ on angle $\omega_i \tau$ and ${\bf
W}_i={\bf W}(\hat {\mbox{\boldmath $\omega$}}_i)$ is a
skewsymmetric matrix (${\bf W}_{\alpha \beta}= -{\bf W}_{\beta
\alpha}$) with ${\bf W}_{XY} =-\hat \omega_Z$, ${\bf W}_{XZ}=\hat
\omega_Y$, ${\bf W}_{YZ} =-\hat \omega_X$ and $\hat
{\mbox{\boldmath $\omega$}}= {\mbox{\boldmath $\omega$}}/\omega$.
Since the decompositions used are correct within an uncertainty of
order ${\cal O}(h^3)$, the trigonometric functions can be replaced
by their rational counterparts (see, e.g., \cite{rotation}), $\cos
\xi =(1-\xi^2/4)/(1+\xi^2/4) +{\cal O}(\xi^3)$ and $\sin \xi =
\xi/(1+\xi^2/4)+{\cal O}(\xi^3)$, which maintain the
orthonormality of ${\bf D}$ and are more efficient for the
computations. Then the spin rotation reduces to
\begin{eqnarray}
{\rm e}^{L_{{\bf s}_i} \tau} {\bf s}_i(t) &=& \Big( {\bf s}_i(t) +
[{\mbox{\boldmath $\omega$}}_i {\mbox{\boldmath $\times$}} {\bf s}_i(t)]
\tau + \frac{\tau^2}{2} \big( {\mbox{\boldmath $\omega$}}_i
({\mbox{\boldmath $\omega$}}_i {\mbox{\boldmath $\cdot$}}
{\bf s}_i(t)) \nonumber \\ &-& \frac12 ({\mbox{\boldmath $\omega$}}_i
{\mbox{\boldmath $\cdot$}} {\mbox{\boldmath $\omega$}}_i)
{\bf s}_i(t) \big) \Big) \Big/ \Big( 1+\Big(\frac{\omega_i \tau}{2}
\Big)^2 \Big) \, .
\end{eqnarray}
This completes the new algorithm.

We note that our basic EOM are time-reversible and exact solutions
behave symplectically. As can be shown, the algorithm derived
reproduces these features, even though the trajectories are
generated with a limited accuracy. Indeed, the initial propagator
was decomposed (Eqs. (3) and (4)) into subparts symmetrically,
and, as a consequence, the final expressions for ${\bf r}_i$,
${\bf v}_i$ and ${\bf s}_i$ will be invariant with respect to the
transformation $h \rightarrow -h$. Furthermore, simple shifts
(applied separately in position and velocity space) do not change
the phase volume. These properties are very important for our
purpose because, as is now well established \cite{Allen,Frenkel},
the stability of an algorithm normally follows from its time
reversibility and symplecticity. Another nice property of the
algorithm is its exact conservation of spin lengths (rotations
given by Eq. (5) do not change the norm of vectors) that is
crucial for the class of models considered.

In our MD study of the Heisenberg fluid, we have used the Yukawa
potential \cite{Nijmei}, $J(r)=(\epsilon \sigma/r)\exp[(\sigma
-r)/\sigma]$, and a soft-core potential \cite{Allen}, $\Phi(r)=
4\varepsilon [(\sigma/r)^{12} -(\sigma/r)^6]+\varepsilon$ at $r <
2^{1/6} \sigma$ and $\Phi(r)=0$ otherwise, for the description of
spin and liquid interactions with the intensities $\epsilon$ and
$\varepsilon$, respectively. The function $J(r)$ was truncated at
$R=2.5\sigma$ and shifted to be zero at the truncation point to
avoid force singularities. The simulations were carried out for
$N=1000$ particles (employing periodic boundary conditions) at a
reduced density $n^\ast=N\sigma^3/V=0.6$, a reduced temperature
$T^\ast=k_{\rm B} T/\epsilon=1.5 < T_{\rm c}^\ast$ (where $T_{\rm
c}^\ast \simeq 2.06$ is the temperature of ferromagnetic
transition \cite{Folk}), a reduced core intensity $\varepsilon
/\epsilon=1$, and a {\em dynamical coupling parameter} $d=\sigma
(m \epsilon)^{1/2}/\hbar=2$. This last parameter presents, in
fact, the ratio $\tau_{\rm tr}/\tau_{s}$, where $\tau_{\rm tr}=
\sigma(m/\epsilon)^{1/2}$ and $\tau_{s}=\hbar/\epsilon$ are the
characteristic time intervals of varying translational and spin
variables, respectively. Since we are investigating a {\em
ferrophase} and dealing with a {\em microcanonical} (NVES)
ensemble, a non-zero magnetization of the system must be specified
additionally. This quantity was taken from our single MC
simulation \cite{Folk}, $\langle {\bf S} \rangle_0/N = 0.6536 \pm
0.0001$, where $\langle \ \rangle_0$ denotes the canonical
averaging. All test runs were started from an identical well
equilibrated configuration. The recalculation of local magnetic
fields (during spin subdynamics (4)) took approximately the same
processor time as that of translational forces, spending in total
0.5 sec per step on the Origin 2000 workstation. It is worth
emphasizing that contrary to spin lattice dynamics \cite{Krech}
(when auxiliary MC cycles are involved to generate equilibrium
configurations as initial conditions for the EOM), the
equilibration of our system can be performed within NVES MD
simulations exclusively (at the specified value for ${\bf S}
\equiv \langle {\bf S} \rangle_0$). This is possible because of
the {\em energy exchange} between the spin and liquid subsystems.

Symmetries of Hamiltonian (1) impose conservation laws on the
total momentum ${\bf P} = m \sum_i {\bf v}_i$, total spin ${\bf
S}=\sum_i {\bf s}_i$ and total energy $E \equiv H$. These three
integrals of motion cannot be conserved perfectly at the same time
within any approximate scheme known. This a typical situation in
MD simulations. The MD results for the total energy $E^\ast=
E/\epsilon$ (subsets (a)-(d)) and total spin $S$ (subsets (e)-(h))
as functions of the length of the simulations are presented in
Fig.~1. Four time steps, namely, $h^\ast= h/\tau_{\rm
tr}=0.00125$, 0.0025, 0.005 and 0.01, were used to integrate the
EOM (solid curves). These results are compared with those obtained
by us using the well established Adams-Bashforth-Moulton (ABM)
predictor-corrector scheme \cite{Burden} (dashed curves in subsets
(a) and (b)). As can be seen from Fig.~1a, the ABM integrator
fulfills energy conservation up to a similar accuracy as our
algorithn at the smallest time step $h^\ast= 0.00125$. However,
for larger step sizes (see Fig.~1b) the ABM scheme is unstable
and, thus, cannot be used. Note that very small step sizes are
impractical because then too much time-consuming force and field
evaluations have to be done during the typical observation times.

No systematic drift in $E(t)$ and $S(t)$ was observed within our
algorithm at time steps up to $h^\ast=0.01$ over a length of
$t/h=$ 100 000. The precision of the algorithm was measured in
terms of the ratio $\Gamma_E= (\langle (E(t)-E(0))^2 \rangle /
\langle (U(t)-U(0))^2 \rangle)^{1/2}$ of total and potential ($U$)
energy fluctuations. Taking into account that for our system
$(\langle U(t)-U(0) \rangle^2)^{1/2} \approx 0.0335$, we have
obtained: $\Gamma_E \approx$ 0.12\%, 0.28\%, 0.98\% and 7.7\% for
the time steps $h^\ast=0.00125$, 0.0025, 0.005 and 0.01,
respectively. In order to reproduce properly the features of
microcanonical ensembles the ratio $\Gamma_E$ should not exceed a
few per cent. As we can see, time steps of $h^\ast \leq 0.01$
satisfy this requirement and, thus, they can be used for precise
calculations.

\begin{figure}[htbp]
\begin{centering}
\begin{picture}(81,178)
\epsfxsize=81mm
\put(0,0){
\framebox{
\epsffile[41 289 590 702]{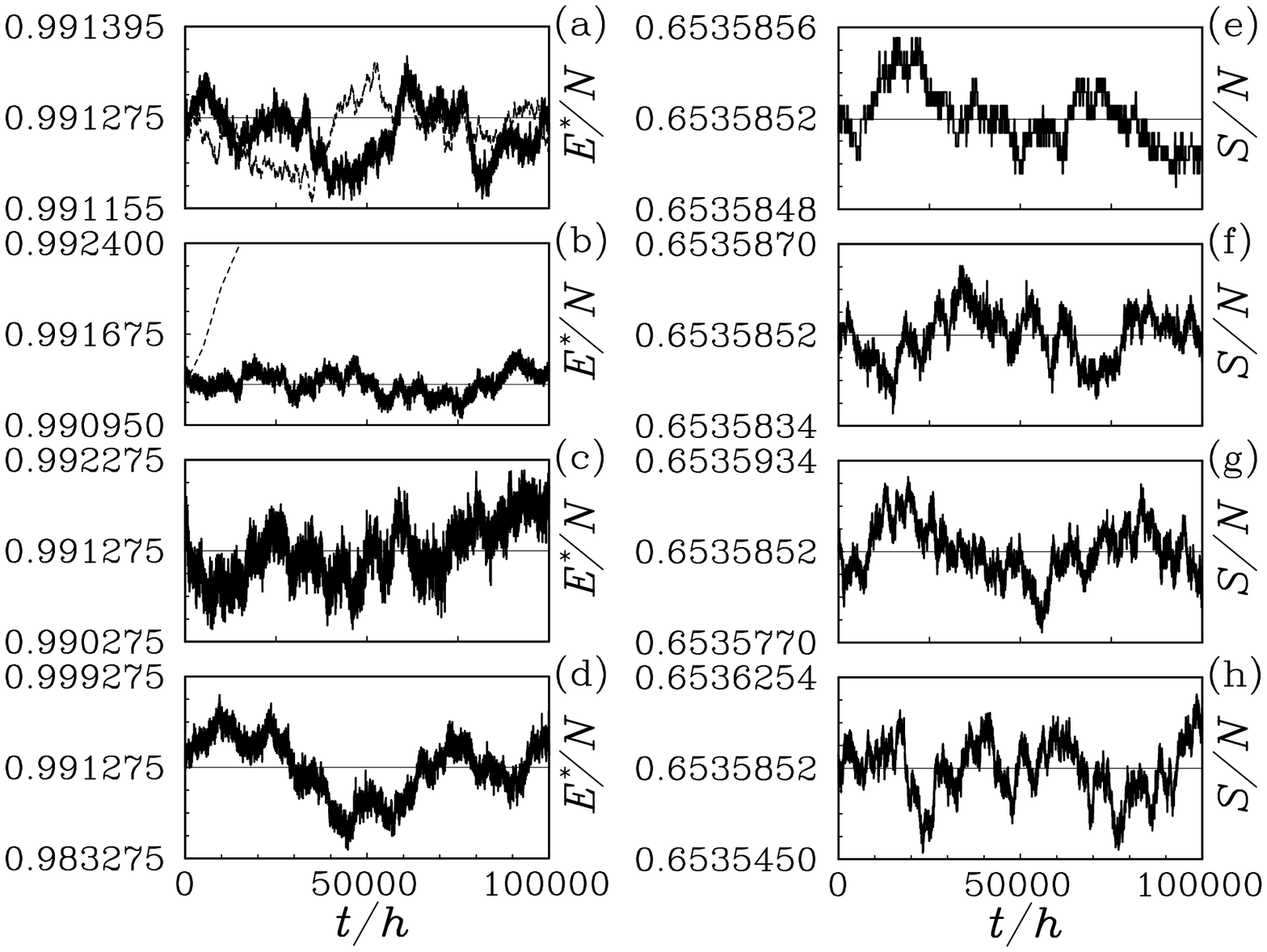}}}
\end{picture}
\end{centering}
\end{figure}
{\small FIG. 1. Reduced total energy $E^\ast(t)/N$
[(a)-(d)] and magnetization $S(t)/N$ [(e)-(h)] per spin as
functions of the observation time obtained within the
decomposition [solid curves] and predictor-corrector [dashed
curves in (a) and (b)] algorithms at four fixed time steps:
$h^\ast=$ 0.00125, 0.0025, 0.005, and 0.01. The values
$E^\ast(0)/N$ and $S(0)/N$ are plotted by the horizontal thin
lines.}

\vspace{18pt}

Note that the decomposition and ABM methods conserve the total
momentum ${\bf P}$ to within machine accuracy. The reason is that
all velocities are updated simultaneously and the interparticle
forces are evaluated exploiting Newton's third law. For similar
reasons, the ABM integration maintains the total magnetization
${\bf S}$ (but it does not conserve spin lengths). In our scheme
the magnetization is not conserved exactly. However, the
fluctuations appear to be very small (see Fig.~1e-1h) and lead to
the values $(\langle ({\bf S}(t)-{\bf S}(0))^2 \rangle)^{1/2}
\simeq 10^{-7}, 5 \cdot 10^{-7}, 2 \cdot 10^{-6}$ and $10^{-5}$ at
$h^\ast=$0.00125, 0.0025, 0.005 and 0.01, respectively.

The spectra $F(k,\omega)=\frac{1}{2\pi} \int_{-\infty}^{\infty}
F(k,t) e^{-{\rm i} \omega t} {\rm d} t$ of the spin-spin $F_{ss}^{\rm L,T}
(k,t)=\langle \sum_{i,j} {\bf s}_i^{\rm L,T}(0) {\mbox{\boldmath $\cdot$}}
\, {\bf s}_j^{\rm L,T}(t) n_{ij}({\bf k},t) \rangle$, density-density
$F_{nn}(k,t)= \langle \sum_{i,j} n_{ij}({\bf k},t) \rangle$ and spin-density
$F_{sn}^{\rm L}(k,t)=\langle \sum_{i,j} {\bf s}_i^{\rm L}(0) n_{ij}({\bf k},
t) \rangle \equiv F_{ns}^{\rm L}(k,t)$ time correlation functions are shown
in Fig.~2. The superscripts (L) and (T) refer to the longitudinal and
transverse components of ${\bf s}_i$ with respect to the vector ${\bf S}$,
and $n_{ij}({\bf k},t)=N^{-1}\exp \{{\rm i}{\bf k} {\mbox{\boldmath $\cdot$}}
({\bf r}_i(0)-{\bf r}_j(t))\}$. These functions were obtained within our
decomposition integration at $h^\ast=0.005$, and the microcanonical
averaging $\langle \ \rangle$ was taken over 100 000 steps for each
of 10 independent runs. A dimensionless representation has been used
for $F^\ast(k,\omega)=F(k,\omega)/\tau_{\rm tr}$ with $\omega^\ast=
\omega \tau_{\rm tr}$, $k^\ast=k/k_{\rm min}$ and $k_{\rm min}=
2\pi/V^{1/3}$.

\begin{figure}[htbp]
\begin{centering}
\begin{picture}(81,116)
\epsfxsize=81mm
\put(0,0){
\framebox{
\epsffile[41 456 563 702]{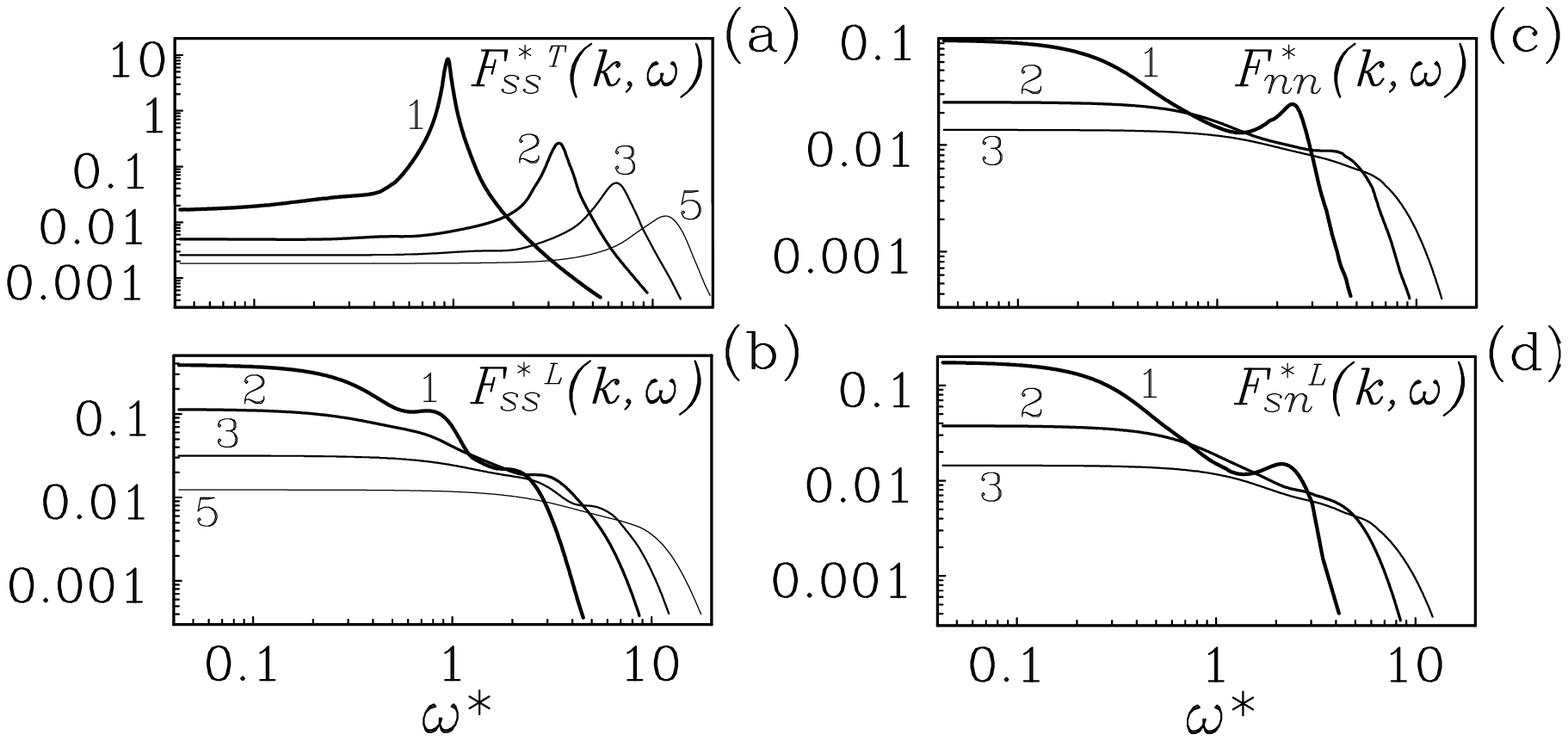}}}
\end{picture}
\end{centering}
\end{figure}
{\small FIG. 2. Transverse (a) and longitudinal (b)
spin-spin, density-density (c), and spin-density (d) functions for
a Heisenberg fluid versus frequency $\omega^\ast$. The curves
corresponding to the wavenumbers $k^\ast= 1,2,3$, and 5 are marked
by "1", "2", "3", and "5", respectively.}

\vspace{18pt}

One peak can be identified for the function $F_{ss}^{\rm
T}(k,\omega)$ at each wavevector $k$. This peak is very sharp at
small $k$ and shifts to the right with increasing $k$. Such a
quasiparticle behavior should be associated with the existence of
{\em transverse spin waves} in the spin liquid. Up to three maxima
were observed for the component $F_{ss}^{\rm L}(k,\omega)$. While
the first maximum at $\omega=0$ corresponds to pure diffusive
processes, the position of the second one coincides with that of
the transverse spin wave peak, indicating the possibility of
propagating {\em longitudinal spin waves} as well which, however,
are damped much stronger. The origin of the third maximum in
$F_{ss}^{\rm L} (k,\omega)$ can be explained by the direct
influence of the liquid subsystem on the spin one, because its
position coincides with a peak position in $F_{nn}(k,\omega)$.
This last peak should be associated with propagative {\em sound
modes} well established for liquid systems \cite{MryglodOmelyan},
whereas a maximum of $F_{nn}(k,\omega)$ at $\omega=0$ represents
the well-known diffusive {\em heat mode}. The function
$F_{sn}^{\rm L}(k,\omega)$ behaves similarly to
$F_{nn}(k,\omega)$. In general the results obtained are in good
agreement with the predictions of Ref.~\cite{MryglodFolk}. The
additional possibility of longitudinal spin wave propagations in
magnetic liquids at sound frequency can be considered as a new
effect which has yet to be observed experimentally. Similar effect
was found also \cite{Folk} in our MD calculations performed for
model (1) at a higher temperature $T> T_{\rm c}$ at the presence
of an external magnetic field. Taking into account the theoretical
results of Ref.~\cite{MryglodFolk} allows us to state that in the
both cases considered the Brillouin sound peaks appear in
$F_{ss}^{\rm L}(k,\omega)$ due to magnetostriction caused by spin
ordering.

In conclusion, we list the chief advantages of the new algorithm
over existing numerical schemes: (i) time-reversibility and
symplecticity; (ii) explicitness (no iteration); (iii) exact
conservation of spin lengths; (iv) much more accuracy in total
energy conservation. Moreover, its excellent stability (allowing
applications with much larger time steps) may lead to a
substantial improvement of the speed of MD simulations for
magnetic liquids. It can also be used for lattices (then only Eqs.
(4) and (5) must be employed) with arbitrary structures. These and
related problems will be considered in a separate publication.

Part of this work was supported by the Fonds zur F\"orderung
der wissenschaftlichen Forschung under Project No. P12422-TPH.

\end{multicols}

\end{document}